\documentclass[twocolumn,twoside,showpacs,showkeys,preprintnumbers,superscriptaddress,amsmath,amssymb,prc,nofootinbib,a4paper]{revtex4}

\usepackage{cases}
\usepackage[normalem]{ulem}
\usepackage{amsmath}
\usepackage{subcaption}
 \usepackage{CJK}
 \usepackage{txfonts}
 \usepackage{mathrsfs}
 \usepackage{amssymb}
 \usepackage{graphicx}
 \usepackage{dcolumn}
 \usepackage{bm}
 \usepackage{floatflt}
 \usepackage{float}
 \usepackage{siunitx}
 \usepackage{rotating}
 \usepackage{tikz}
\usepackage{physics}
 \usepackage{multirow}
 \usepackage[bookmarksnumbered,bookmarksopen,colorlinks,citecolor=blue,linkcolor=blue]{hyperref} %
 \usepackage{epsfig,graphics,psfrag}
\usepackage{xcolor}

\def\esym{$E_{\rm sym}(\rho)$}
\def\thu{Department of Physics, Tsinghua University, Beijing 100084, China}
\def\riken{RIKEN Nishina Center, Hirosawa 2-1, Wako, Saitama 351-0198, Japan}
\def\gxnu{College of Physics and Technology, Guangxi Normal University, Guilin 541004, China}
\def\gxkl{Guangxi Key Laboratory of Nuclear Physics and Technology, Guangxi Normal University, Guilin 541004, China}

\begin{document}
      
 \title{Simulation studies of the isovector reorientation effect of deuteron scattering on heavy target}

 \author{Baiting Tian}
 \email[Corresponding author: ]{tbt23@mails.tsinghua.edu.cn}
 \affiliation{\thu}

 \author{Boyuan Zhang}
 \affiliation{\thu}

\author{Dawei Si}
\affiliation{\thu} 

\author{Sheng Xiao}
\affiliation{\thu} 

\author{Yijie Wang}
\email[Yijie Wang:]{yj-wang15@tsinghua.org.cn}
\affiliation{\thu}

\author{Tadaaki Isobe}
\email[Tadaaki Isobe:]{isobe@riken.jp}
\affiliation{\riken}

\author{Hideaki Otsu}
\email[Hideaki Otsu:]{otsu@ribf.riken.jp}
\affiliation{\riken}

\author{Li Ou}
\email[Li Ou:]{liou@gxnu.edu.cn}
 \affiliation{\gxnu}
 \affiliation{\gxkl}

\author{Zhigang Xiao}
\email[Zhigang Xiao:]{xiaozg@tsinghua.edu.cn}
\affiliation{\thu}

\date{\today}


\begin{abstract}

The isovector reorientation (IVR) effect of deuteron scattering on heavy target provides a novel means to probe the nuclear isovector potential, which gives rise to the nuclear symmetry energy.  The simulation studies on the experimental measurement of IVR  effect using the SAMURAI terminal at RIKEN Nishina center have been performed to demonstrate the feasibility of the experiment. By introducing a well-designed polarimeter to detect the $\mathrm{p}(\vec{\mathrm{d}}, \mathrm{d})\mathrm{p}$ elastic scattering,  monitoring of the tensor polarization of the deuteron beam can be implemented. The protons and neutrons produced by the breakup of polarized deuterons scattering off heavy targets are designed to be measured by proton drift chamber (PDC) combined with the SAMURAI magnet and NEBULA \textcolor{blue}(Neutron-detection system for Breakup of Unstable-Nuclei with Large Acceptance) detector, respectively. The detector responses are simulated using \texttt{GEANT4} framework, where the events of the deuteron elastic breakup are generated by an Improved Quantum Molecular Dynamics model. The results of reconstructing the deuteron breakup events demonstrate the feasibility of detecting the IVR effect at SAMURAI with both longitudinal and transverse tensor polarized deuteron beams with a polarization degree of approximately 80\%.

\end{abstract}


 \maketitle
 
\section{Introduction}

The symmetry energy characterizes the isospin-dependent part of the nuclear equation of state, quantifying the energy cost associated with neutron-proton asymmetry \cite{li2008recent,danielewicz2002determination,Li2021}. The density dependence of the symmetry energy, \esym, has drawn increasing attention from both nuclear physics and astrophysics communities. However, the stiffness of \esym~ still suffers from significant uncertainty, particularly at supra-saturation densities \cite{Li:2019xxz}. Recently, enormous progress has been made from heavy ion experiments \cite{Tsang:2004zz,SpiRIT:2021gtq,Xiao:2008vm,Zhang:2017xtk} and observations of gravitational waves generated by neutron star merger \cite{LIGOScientific:2017vwq,LIGOScientific:2018cki}, as well as the combination of the two \cite{Tsang:2023vhh,Huth:2021bsp}. Further elaborations are still ongoing in searching new probes of \esym \cite{Wang:2021mrv,Wang:2022ysq,Wang:2023kyp}.

Using Hugenholtz–Van Hove theorem \cite{Hugenholtz:1958zz}, the relationship between symmetry energy and single particle potential ($U_{\mathrm{n/p}} = U_0 \pm U_{\mathrm{sym}}(\rho, k_{\mathrm{F}})$) can be easily derived. We can decouple the symmetry energy into several parts related to the single-particle potential and Fermi kinetic energy \cite{PhysRevC.82.054607} 
\begin{align}\label{Esym2}
E_{\rm sym}(\rho) &= \frac{1}{6} \frac{\partial
[t(k)+U_0(\rho,k)]}{\partial k}\Big|_{k_{\rm F}} \cdot k_{\rm F} + \frac{1}{2}
U_{\rm sym}(\rho,k_{\rm F})\nonumber \\ 
&=\frac{1}{3} t(k_{\rm F}) + \frac{1}{6}
\frac{\partial U_0}{\partial k}\Big|_{k_{\rm F}}\cdot k_{\rm F} +
\frac{1}{2}U_{\rm sym}(\rho,k_{\rm F})  \\
&= \frac{k_{\rm F}^2}{6m^*_0 (\rho,k_{\rm F})} + \frac{1}{2} U_{\rm sym}(\rho,k_{\rm F}), \nonumber
\end{align}
Here, $t(k_{\mathrm{F}})$ is the nucleon kinetic energy at the Fermi momentum, and $k_{\mathrm{F}} = \left( \frac{3\pi^2\rho}{2} \right)^{1/3}$ is the Fermi momentum for symmetric nuclear matter with density $\rho$. Equation~\ref{Esym2} shows that the microscopic origin of the symmetry energy is the difference between the neutron and proton Fermi surfaces. Equation \ref{Esym2} shows that the microscopic origin of the symmetry energy is the difference between the neutron and proton Fermi surfaces. For the isoscalar potential $U_0(\rho, k)$, information on its density and momentum dependence has been obtained from high-energy heavy-ion collisions (e.g., see Ref. \cite{danielewicz2002determination}). While existing data constrain its behavior to some extent, significant uncertainties persist in high-momentum and high-density regimes. In contrast, the isovector potential $U_{\mathrm{sym}}(\rho, k)$, particularly at extreme densities and momenta, remains poorly constrained. This potential has been identified as a critical parameter for resolving ambiguities in the high-density behavior of the symmetry energy \cite{xu2011analytical,chen2012single,li2008recent}. At saturation density, the isoscalar effective mass $\frac{m_0^*}{m} = 0.70\pm0.05$, and the isovector potential $U_{\rm sym}(\rho_0,k_{\rm F})=28.7 \pm 7.8~\mathrm{MeV}$ \cite{li2013constraining,li2015neutron}. Therefore, the challenge of excluding the interference of isoscalar forces in heavy-ion experiments while extracting the characteristics of isovector forces has become critical.
  
\begin{figure}
    \centering
    \includegraphics[width=0.45\textwidth]{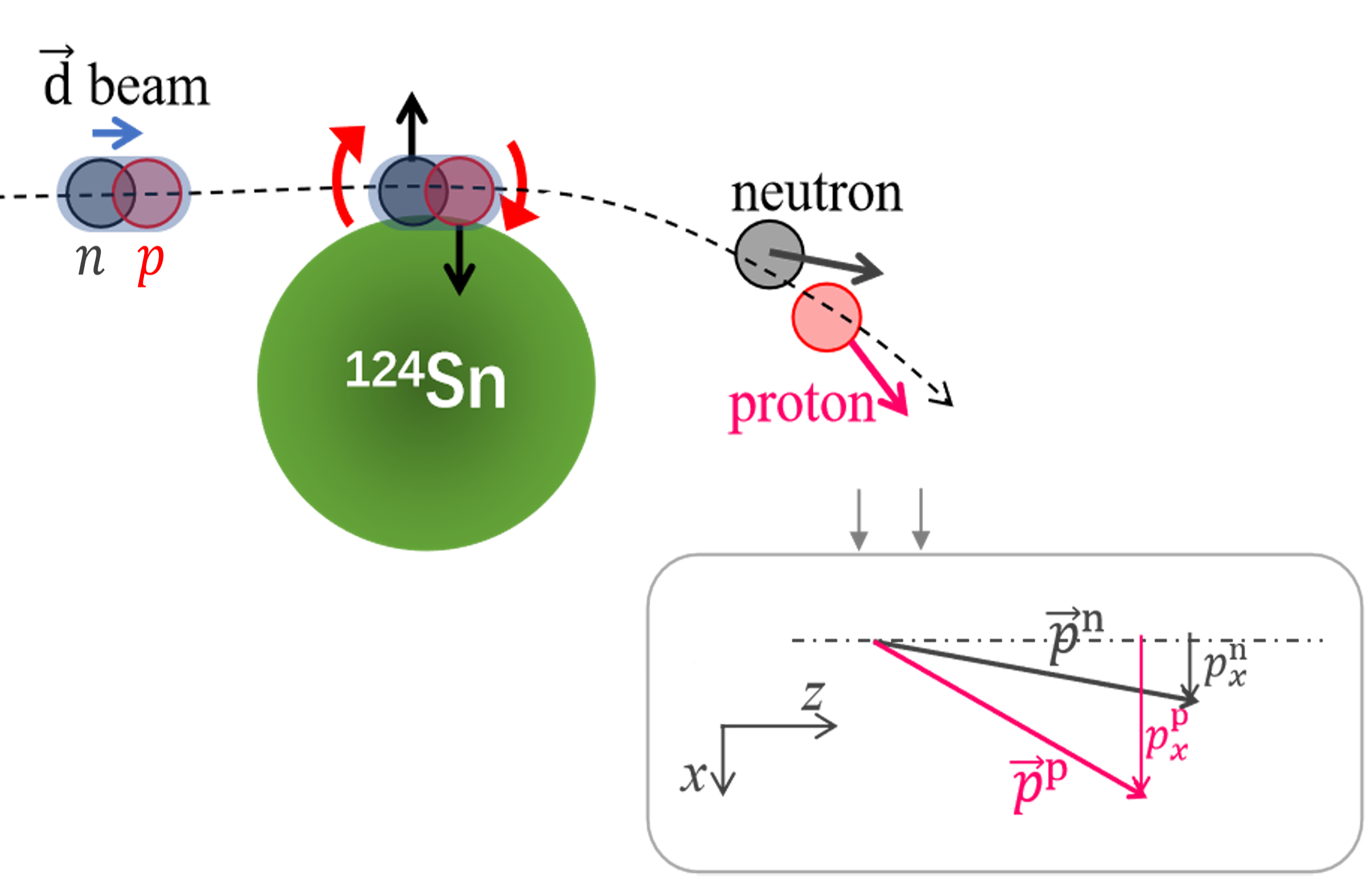}
    \caption{The schematic view of the isovector reorientation effect of a polarized deuteron scattering on a $^{124}$Sn target, followed by the elastic breakup. The inset presents the momentum vectors to be used in the following analysis.}
    \label{fig:ivr-view}
\end{figure}

The isovector reorientation (IVR) effect effectively mitigates the influence of the isoscalar potential, enabling the extraction of isovector information with high sensitivity \cite{PhysRevLett.115.212501,PhysRevC.101.024603}. As shown in Figure \ref{fig:ivr-view}, in the peripheral scattering of a polarized deuteron beam on a heavy target, the isovector force, which is attractive to protons and repulsive to neutrons, acts like a torque on the deuteron, inducing an additional rotation of the latter. Consequently, the subsequent breakup incorporates the rotational effect, which exhibits high sensitivity to the isovector force. Detailed simulations demonstrate that the angular
distribution of the correlated neutron and proton from the deuteron breakup is highly sensitive to the isovector potential, yet insensitive to the isoscalar nuclear potential. The experimental observable is straightforward to implement: measuring the angular distribution of the relative momentum vector between neutrons and protons arising from the breakup of a polarized deuteron. This observable acts as a probe with exceptional sensitivity to the symmetry energy in regions of low nuclear density. By employing this probe, the uncertainties in the density dependence of the symmetry energy can be effectively constrained.

Among various research and development (R\&D) works, two key issues have to be addressed before conducting the experiment at SAMURAI, RIKEN Nishina center.  One is the monitoring of the polarized deuteron beam, the other is the design (and optimization) of the detector setup to measure the breakup of the polarized deuteron scattering on heavy target, which carries the IVR effect. These two issues can be elucidated by simulation studies, which make the current paper. The remainder of this paper is organized as following. In Section \ref{sec:exp_plan}, we outline the experimental plan and describe the simulation methods employed in this study. Section \ref{sec:MonitorPol} focuses on the monitoring of the polarized beam, where we define tensor polarization and introduce two monitoring options. Moreover,   the design of the polarimeter and  the simulation results for beam monitoring are presented. Section \ref{model_desc} describes the Improved Quantum Molecular Dynamics(ImQMD) transport model as the event generator in our simulations. Section \ref{sec:AngDistribution} present the simulation of the detector response at SAMURAI and the event reconstruction. We detail the detector setup, target position optimization, and the responses of both neutrons and protons, along with the background simulation and the beam dumper. Additionally, we observe the IVR effect within the current experimental design. Finally, Section \ref{sec:Conclusion} provides the conclusion and outlooks of the study.

\section{A brief introduction to the beam experiment}
\label{sec:exp_plan}

\begin{figure*}
    \centering
    \includegraphics[width=\textwidth]{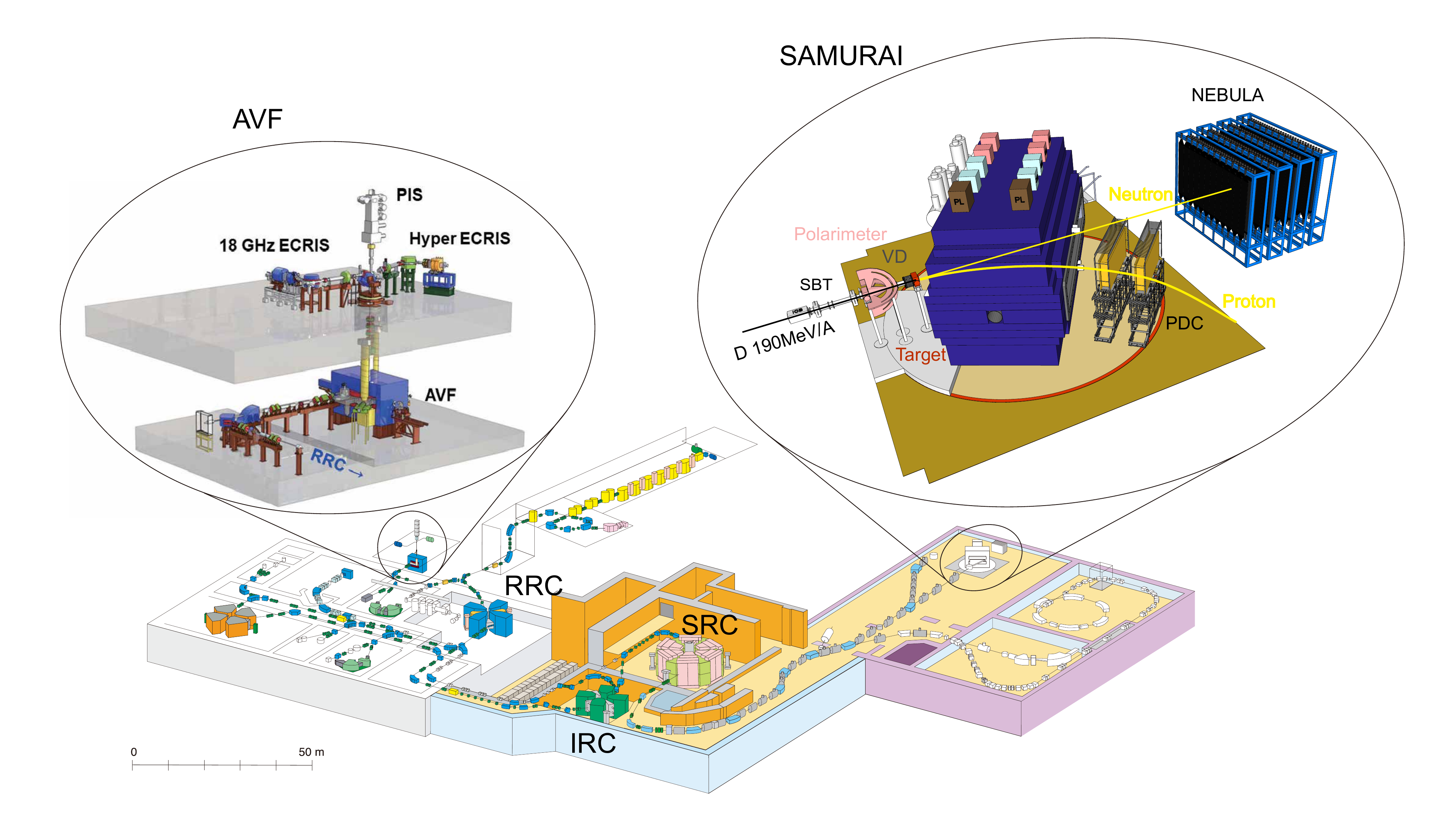}
    \caption{The bird's-eye view of the experiment. The AVF (Azimuthally Varying Field Ring Cyclotron), RRC (Riken Ring Cyclotron) and SRC (Superconducting Ring Cyclotron) are used to produce, accelerate and transport the polarized deuteron beam to 190 MeV/u. Beam trigger (SBT) is to provide the ount number of
beam and determine start timing to calculate the time of flight(TOF). The Polarimeter monitors the tensor polarization of the deuteron beam before it reaches the target. The target is positioned at the entrance of the SAMURAI spectrometer, which is a large-acceptance superconducting dipole magnet for momentum analysis of charged particles. The PDC (Proton Drift Chamber) detects breakup protons downstream of the target, while NEBULA is used to detect neutrons.The VD (veto detector) is to veto the central collision based on the multiplicity.  }
    \label{fig:exp_plan}
\end{figure*}

The experimental setup is specifically optimized to measure the isovector reorientation effect in deuteron scattering on heavy targets. Figure \ref{fig:exp_plan} presents the overview of the beam line and detector arrangement, highlighting the strategic placement of key components to maximize detection efficiency and minimize background. The deuteron beam is polarized and accelerated to 190 MeV/u. The tensor polarized beam is provided by the polarized ion source and accelerated by the AVF (Azimuthally Varying Field Ring Cyclotron) and RRC (Riken Ring Cyclotron) successively, and then by SRC (Superconducting Ring Cyclotron) through the IRC (Intermediate-stage Ring Cyclotron) bypass transport beam line. The spin symmetry axis of the deuteron beams is controlled by the Wien Filter prior to acceleration. Single-turn extractions are required for the three cyclotrons to maintain polarization amplitudes during acceleration. The polarization of the deuteron beam
is typically 80\% of the theoretical value \cite{sakamoto2016acceleration}.

    The experimental IVR effect is detectable  in longitudinal ({$p_{zz}$}) and vertical ({$p_{yy}$}) polarization conditions. The deuteron-induced scattering involves both nuclear and Coulomb interactions. To differentiate nuclear and Coulomb effects, it is planned to use three different targets, $^{112}\rm{Sn}$, $^{124}\rm{Sn}$, and $^{208}\rm{Pb}$. The  $^{124}\rm{Sn}$ and the $^{208}\rm{Pb}$ have the same neutron abundance ($N/A$ = 0.60 and 0.61, respectively) but different atomic charge,
therefore the isospin effect is alike and the difference is expected due to the Coulomb force.

The experimental setup is specifically optimized to mea-
sure the isovector reorientation effect in deuteron scattering on
heavy targets.  According to the Improved Quantum Molecular Dynamics (ImQMD) model
calculations, for the scattering events with impact parameter $b=7\sim9\,\mathrm{fm}$
which is the range showing the most sensitivity of IVR effect on the stiffness of the symmetry energy $E_{\rm sym}(\rho)$, as characterized by the parameter of  $\gamma$ in Eq. (\ref{Esymwithgamma}), the most probable direction of the breakup neutrons is not very close to $\theta=0{^\circ}$.
Considering that NEBULA is difficult to move, in order to measure
the neutrons in the most favorable range of $\theta_{\mathrm{n}}$,
it is suggested to place the target upstream of the magnet, at a distance of 3 meters from the center of the magnet, which ensures that the range of $-10{^\circ}\le\theta_{\mathrm{n}}\le10{^\circ}$ is covered for the most probable neutron emissions.

As for the SAMURAI magnet, we  choose $B=1.2\,\mathrm{T}$ and
the $30{^\circ}$ configuration, which fits mostly our experimental
purpose. In addition, we do not require the magnet chamber to be in
vacuum, since the beam trajectory is influenced little by the air
at the current beam energy. By removing the downstream exit window of the SAMURAI chamber, we
have a wider range of exit angle for both neutrons and protons. 
According
to the simulation results, the reorientation effect can be observed
for both $z$-polarization ($p_{zz}$) and $y$-polarization ($p_{yy}$). Therefore,
we prefer to use either the $z$-polarization or $y$-polarization in our
experiment, depending on the machine condition. The unreacted deuteron beam will be collected
by the beam dumper, which is surrounded by a water tank to shield the neutron background radiation.

To achieve the experimental goal, the detector system is arranged with a polarimeter and a veto detector upstream of the target to monitor deuteron polarization and suppress inelastic background, respectively. Downstream of the target, protons from deuteron breakup are detected by the proton drift chamber (PDC), while neutrons are detected by the NEBULA array. The PDC is positioned 3 meters from the target, and NEBULA is placed 7.3 meters from the target, ensuring optimal detection efficiency for both protons and neutrons.

\section{Monitor the tensor polarization }
\label{sec:MonitorPol}

The design with a polarized deuteron beam is essential in this experiment.
If the deuteron beam were unpolarized, the random orientation would average out the differences in the proton-neutron angular distribution, making it impossible to observe the IVR effect.
We monitor the deuteron beam's polarization using a method based on d-p elastic scattering. This involves measuring the cross-sections at different angles to confirm the polarization state. While the theoretical framework for polarized nuclear reactions is well established and experimental techniques for monitoring $p_{y'y'}$ are available, the direct measurement of $p_{z'z'}$ remains unexplored. This work addresses this gap by developing a robust monitoring mechanism for $p_{z'z'}$. To avoid misunderstandings caused by different notational conventions, we start from the most fundamental principles and derive our polarization monitoring mechanism in detail.

\subsection{Definition of  the  tensor  polarization}

Deuteron is a spin-1 nucleus.
 The polarization state of the deuteron can be fully described using its spin density operator.
 This operator characterizes a statistical mixture of pure quantum states,
\begin{align}
\hat{\rho} = \sum_i Q_i \ket{\phi_i}\bra{\phi_i},
\end{align}
where $\hat{\rho}$ is the density operator (distinct from the nuclear matter density $\rho$ in Eq.~\ref{Esym2}) and $Q_i$ is the probability of the pure state $\ket{\phi_i}$.
 For a spin-1 system such as the deuteron, the density matrix $\rho$ is a $3 \times 3$ Hermitian matrix that encodes the information about both the vector and the tensor polarization, the latter of which is required for this experiment.

Here we employ the Cartesian basis for the expansion, consistent with the convention in reference \cite{ohlsen1972polarization}, as it aligns with the definition of the tensor polarization of the deuteron beam used in the experiment along the z and y directions. Although Cartesian tensors suffer from the reducibility difficulty, i.e., they can be decomposed into three independent components with different transformation properties under rotation corresponding to angular momentum multiplicities $L=0$, 1, and 2, respectively. Cartesian tensors can be decomposed into three tensors that transform according to the spherical harmonics of order 0, 1, and 2. Therefore, spherical tensors are more fundamental and better preserve the symmetry. Cartesian and spherical bases can be readily transformed into each other, as detailed in references \cite{ohlsen1972polarization} and \cite{rose1995elementary}.  Specifically, for the deuteron with spin $S=1$, the vector polarization is given by $\mathscr{P}_i = S_i$, and the tensor operator is expressed as $\mathscr{P}_{ij}= 3 S_i S_j - 2I$, where $S_i$ and $S_j$ are the spin operators and $I$ is the identity matrix. Consequently, the density matrix can be expanded in terms of these operators.
\begin{align}
     \hat{\rho} = \frac{1}{3} \Big\{
    I + \frac{3}{2} \left( p_x \mathscr{P}_x + p_y \mathscr{P}_y + p_z \mathscr{P}_z \right)  \nonumber\\  + 
     \frac{2}{3} 
    \left( 
    p_{xy} 
    \mathscr{P}_{xy} + 
    p_{yz} \mathscr{P}_{yz} + p_{xz} \mathscr{P}_{xz} 
    \right) 
 \nonumber\\ 
    + 
    \frac{1}{3} \left( p_{xx} \mathscr{P}_{xx} + p_{yy} \mathscr{P}_{yy} +p_{zz} \mathscr{P}_{zz} \right)
    \Big\},
\end{align}
Where the terms $p_i$ and $p_{ij}$ represent the vector and tensor polarization components of the initial deuteron beam, respectively.

However, $\mathscr{P}_{x x}$,$\mathscr{P}_{y y}$,$\mathscr{P}_{z z}$ is linear dependent. Based on $\mathscr{P}_{x x} + \mathscr{P}_{y y} + \mathscr{P}_{z z} = 0$, we can rewrite the density matrix as other forms.

\subsection{Monitor method}

In quantum mechanics, the final state and the initial state can be connected by the scattering matrix $M$,
    \begin{align}
        \ket*{\phi_{\mathrm{f}}}  = M \ket*{\phi_{\mathrm{i}}},
    \end{align}

So, the relationship between final system's density matrix and the initial system's density matrix reads
    \begin{align}
        	\hat{\rho}_{\mathrm{f}} &= \sum {Q}_i \ket*{\phi_{\mathrm{f}}} \bra*{\phi_{\mathrm{f}}} = \sum {Q}_i M \ket*{\phi_{\mathrm{i}}} \bra*{\phi_{\mathrm{i}}} M^\dagger  \nonumber \\
        &=M\hat{\rho}_{\mathrm{i}} M^\dagger.
    \end{align}

    Defining the analysing power as 
\begin{align}
    A_i = M \mathscr{P}_i M^\dagger ,\\
    A_{ij} = \frac{{\rm Tr} (M \mathscr{P}_{ij} M^\dagger)}{{\rm Tr}(MM^\dagger)}.
\end{align}

One can write the relationship between unpolarized and polarized differential scattering cross-section as

\begin{align}
    \sigma&= \frac{1}{3} \left\{ MM^\dagger + MM^\dagger \frac{3}{2} \sum p_i A_i   + MM^\dagger \frac{1}{3}\sum_{i \neq j} p_{ij}A_{ij}  \right\} \nonumber\\ 
    &= \sigma_0(\theta) \left\{ 1 + \frac{3}{2} \sum_i p_i A_i  + \frac{1}{3} \sum_{ij} p_{ij} A_{ij}\right\} 
\label{primary_pol_crosssection},
\end{align}
where $\sigma_0(\theta) =\frac{1}{3} {\rm Tr} M \hat{\rho}_{\rm unpol} M^\dagger =\frac{1}{3} {\rm Tr} M M^\dagger$.  Here, $\sigma$ is the differential cross-section for a polarized beam, $M$ is the scattering matrix describing the reaction, and $M^\dagger$ is its Hermitian conjugate. The terms $p_i$ and $p_{ij}$ represent the vector and tensor polarization components of the initial deuteron beam. Here, $p$ corresponds to the statistical probability of the polarization operator $\mathscr{P}$, which is equivalent to the $Q$ defined earlier. $A_i$ and $A_{ij}$ are the corresponding vector and tensor analyzing powers, which characterize the sensitivity of the reaction to the polarization. $\sigma_0(\theta)$ is the unpolarized differential cross-section at scattering angle $\theta$. This formula shows how the measured cross-section depends on both the polarization state of the beam and the analyzing powers of the reaction.

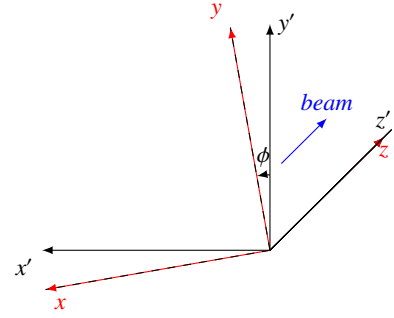
\begin{figure}
\begin{center}
\begin{tikzpicture}[>=latex, x={(1cm,0cm)}, y={(0cm,1cm)}, z={(0.5cm,0.5cm)}]
\draw[->] (0,0,0) -- (-3,0,0) node[below left] {$x'$};
\draw[->] (0,0,0) -- (0,3,0) node[right] {$y'$};
\draw[->] (0,0,0) -- (0,0,3) node[above] {$z'$};
\draw[->,red] (0,0,0) -- (-{3*cos(10)}, -{3*sin(10)}, 0) node[below right] {$x$};
\draw[->,red] (0,0,0) -- ({-3*sin(10)}, {3*cos(10)}, 0) node[above left] {$y$};
\draw[->,red] (0,0,0) -- (0,0,3) node[below] {$z$};
\draw[->,blue] (0,1,0.3) -- (0,1,1.5) node[above]{$beam$};
\draw (0,0,0) -- (0.8,0.8,0) -- (0.8,0.8,1.6) -- cycle;
\draw[->] (0,1,0) arc (90:100:1) node[midway,above] {$\phi$};
\filldraw (0,0,0) circle (0.08) node[below right] {O};
\end{tikzpicture}
\end{center}
\caption{Reference frame. The red axes represent the projectile helicity frame, with the $xoz$ plane defined as the reaction plane.}
\label{fig:frame}
\end{figure}
 Figure \ref{fig:frame} describes the coordinate system applied here. The S coordinate system is the projectile helicity frame, where its z-axis aligns with the direction of the incident particle $k_{\text{in}}$, and the y-axis is perpendicular to both the incident particle direction and the outgoing particle direction, which is along $k_{\text{in}} \times k_{\text{out}}$. In the beam frame $S'$, the $z'$ axis aligns with the incident particle direction $k_{\text{in}}$, and the $y'$ axis points upwards. In Figure \ref{fig:frame}, $\phi$ represents the azimuthal angle. The motion direction of the outgoing particle in $S'$ is given by $(\sin \theta \sin \phi, \sin \theta \cos \phi, \cos \theta)$.
In the S frame(projectile helicity frame), we analyze the properties of the analyzing power $A_i$. In S coordinate system, $A_i$ depends only on $\theta$ and not on $\phi$, which makes our analysis convenient. And in the $S'$ system, it is very convenient to represent the polarization state of the deuteron in the experiment. For example, the two ideal states we need are $p_{z'z'}=1$ and $p_{y'y'}=1$.

To simplify the expression and reduce redundant parameters in Eq.~(\ref{primary_pol_crosssection}), it is essential to consider the symmetries of the system. Specifically, this can be achieved by applying coordinate transformations and utilizing parity conservation. So, it should be noted that in Eq.~\ref{primary_pol_crosssection}, a coordinate transformation is required so that $p$ is defined in the $S'$ frame and $A$ is defined in the $S$ frame. Applying the coordinate transformation for first-order and second-order tensors, one obtains $\overrightarrow{p} = U \overrightarrow{p'}$ and $\overleftrightarrow{p} = U \overleftrightarrow{p'} U^\dagger$, respectively. Under parity transformation, except for pseudovectors like spin, all vectors reverse direction. Therefore, $k_{\text{in}} \rightarrow -k_{\text{in}}$, $k_{\text{out}} \rightarrow -k_{\text{out}}$, and $k_{\text{in}} \times k_{\text{out}} \rightarrow k_{\text{in}} \times k_{\text{out}}$. Since $x$ and $z$ are linear combinations of $k_{\text{in}}$ and $k_{\text{out}}$, they change sign under parity transformation, while the $y$-axis remains unchanged. If parity is conserved, the transformed system should be identical to the original system, implying that all coefficients, including those related to the analyzing power, remain the same before and after the transformation.
Parity conservation requires that an observable must have an even $N_x + N_z$ for non-zero values, where $N_x$ and $N_z$ represent the number of occurrences of $x$ and $z$ as subscripts, respectively. The subscripts represent the components of a tensor. When $N_x + N_z$ is odd, the components of the tensor change sign under coordinate transformation because of parity, but due to the restriction from parity conservation, the observable can only be zero. Thus, we obtain equation \ref{polar_crosssection}.
    \begin{align}\label{polar_crosssection}
    \frac{\sigma(\theta, \phi)}{\sigma_0(\theta)}=1 & +\frac{3}{2}\left(p_{x^{\prime}} \sin \phi+p_{y^{\prime}} \cos \phi\right) A_y(\theta) \nonumber \\
    &+\frac{2}{3}\left(p_{x^{\prime} z^{\prime}} \cos \phi-p_{y^{\prime} z^{\prime}} \sin \phi\right) A_{x z}(\theta)  \nonumber \\
    &+\frac{1}{6}\left[\left(p_{x^{\prime} x^{\prime}}-p_{y^{\prime} y^{\prime}}\right) \cos 2 \phi-2 p_{x^{\prime} y^{\prime}} \sin 2 \phi\right] \nonumber \\  &\left[A_{x x}(\theta)-A_{yy}(\theta)\right] +\frac{1}{2} p_{z^{\prime} z^{\prime}} A_{z z}(\theta) .
    \end{align}
So we can monitor the polarization state of the deuterium beam by placing detectors at the same polar angle $\theta$ but different azimuthal angles at $\phi=0^\circ,~90^\circ,~180^\circ$ and $270^\circ$, marked by L, U, R and D, respectively.  This allows us to measure the components $p_{x'}$, $p_{y'}$, $p_{x'z'}$, $p_{y'z'}$, $p_{z'z'}$, $p_{x'y'}$, and $p_{x'x'}-p_{y'y'}$ of the polarization state as
\begin{widetext}
	\begin{equation}
		\begin{aligned}
			\sigma_L&=\sigma_0\left\{1+\frac{3}{2} p_{y^{\prime}} A_y+\frac{2}{3} p_{x^{\prime} z^{\prime}} A_{x z}+\frac{1}{6}\left(p_{x^{\prime} x^{\prime}}-p_{y^{\prime} y^{\prime}}\right)\left(A_{x x}-A_{y y}\right)+\frac{1}{2} p_{z^{\prime} z^{\prime}} A_{z z}\right\}, 
			\\ 
			\sigma_R&=\sigma_0\left\{1-\frac{3}{2} p_{y^{\prime}} A_y-\frac{2}{3} p_{x^{\prime} z^{\prime}} A_{x z}+\frac{1}{6}\left(p_{x^{\prime} x^{\prime}}-p_{y^{\prime} y^{\prime}}\right)\left(A_{x x}-A_{y y}\right) +  \frac{1}{2} p_{z^{\prime} z^{\prime}} A_{z z}\right\}, 
			\\ 
			\sigma_U&=\sigma_0\left\{1-\frac{3}{2} p_{x^{\prime}} A_y+\frac{2}{3} p_{y^{\prime} z^{\prime}} A_{x z}-\frac{1}{6}\left(p_{x^{\prime} x^{\prime}}-p_{y^{\prime} y^{\prime}}\right)\left(A_{x x}-A_{y y}\right)+  \frac{1}{2} p_{z^{\prime} z^{\prime}} A_{z z}\right\} ,
			\\ 
			\sigma_D&=\sigma_0\left\{1+\frac{3}{2} p_{x^{\prime}} A_y-\frac{2}{3} p_{y^{\prime} z^{\prime}} A_{x z}-\frac{1}{6}\left(p_{x^{\prime} x^{\prime}}-p_{y^{\prime} y^{\prime}}\right)\left(A_{x x}-A_{y y}\right)+  \frac{1}{2} p_{z^{\prime} z^{\prime}} A_{z z}\right\}.
		\end{aligned} 
		\label{pzz-eqs}
	\end{equation}
\end{widetext}

    In paper \cite{bieber2001performance} the method to monitor $p_{y'y'}$ is described. Note that their axis convention for the deuteron polarization differs from the convention used here, and their Eq (5) assumes the polarization axis along $y'$. Using the notation adopted in this work, one first defines the measurable count asymmetry

\begin{align}
\begin{aligned}
     R_{LRUD} &= \frac{N_L+N_R-N_U-N_D}{N_L+N_R+N_U+N_D}= \frac{p_{y'y'}(A_{xx}-A_{yy})}{2p_{y'y'}A_{zz}-4}, \\ 
     p_{y'y'} &= \frac{R_{LRUD}}{\frac{1}{2} A_{zz} R_{LRUD}-\frac{1}{4}(A_{xx}-A_{yy})} .\label{pyy} 
    \end{aligned}
\end{align}

For the monitoring of $p_{z'z'}$, we define the average differential cross section of the four detectors as $\Bar{\sigma} = \frac{\sigma_L+\sigma_R+\sigma_D+\sigma_U}{4}$. The relationship between the average intensity and the tensor polarization is given by Eq. (\ref{pzz_average_detector}). If we know the analyzing power, we can determine the tensor polarization from the average intensity. Alternatively, we can take measurements at two different angles to eliminate the effects of beam intensity and other factors:

    \begin{align}
       \Bar{\sigma}  \equiv \frac{\sigma_L+\sigma_R+\sigma_D+\sigma_U}{4} = \sigma_0 \left(1+\frac{1}{2}p_{z'z'}A_{zz}\right). \label{pzz_average_detector}
    \end{align}

\subsection{The data of the Analysing Power and the Simulations}

\begin{figure}[htbp]
    \centering
    \includegraphics[width=0.45\textwidth]{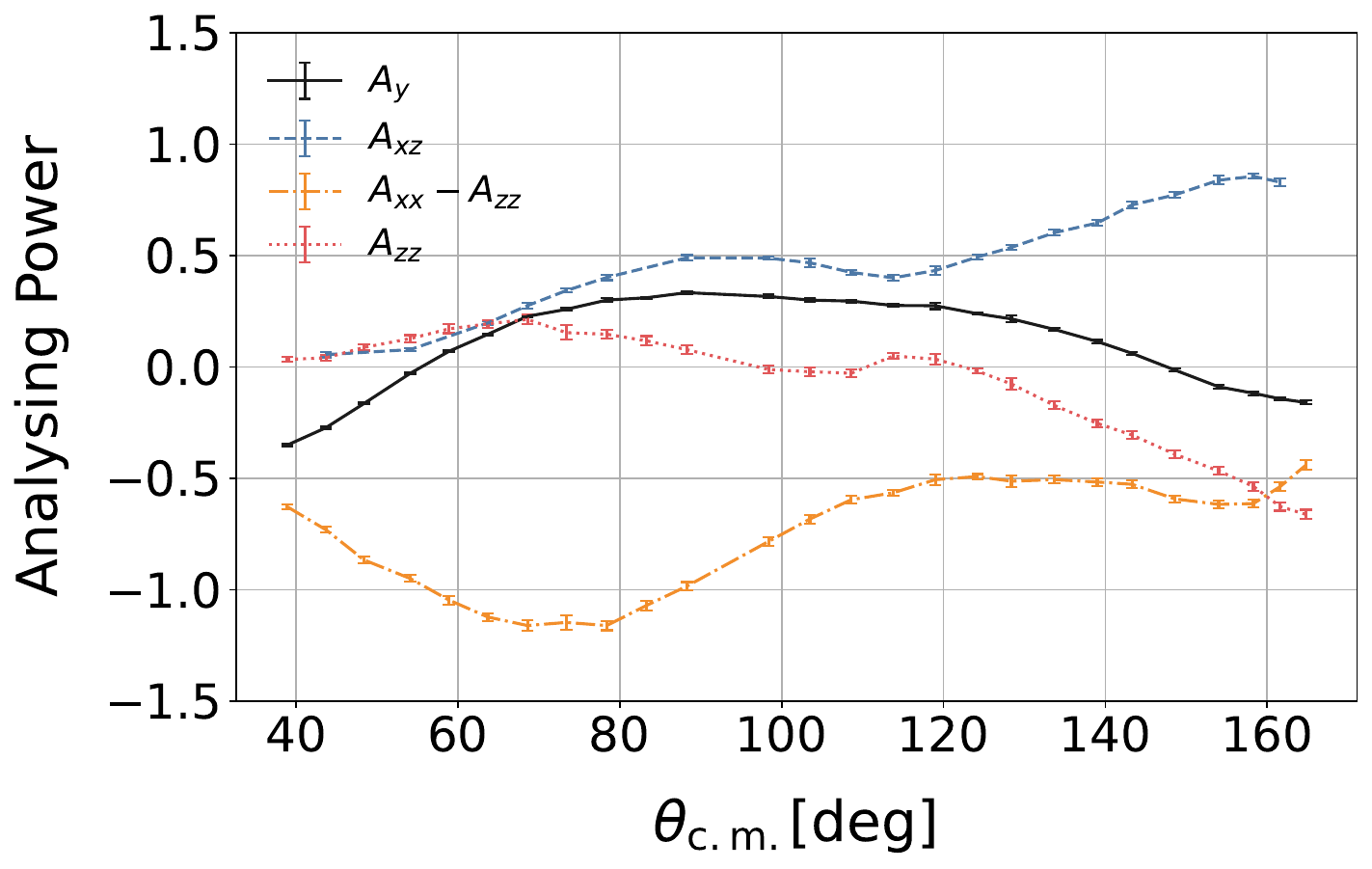}
    \caption{The relationship between the analyzing power and the center-of-mass scattering angle $\theta_{\mathrm{c.m.}}$, data taken from \cite{SekiguchiK}.}
    \label{analysing_power}
\end{figure}

    In the paper \cite{SekiguchiK},  the analyzing power data for the deutron proton elastic scatering, D(p,EL)D process, is plotted in Figure \ref{analysing_power}.

\begin{figure}
    \centering
    \includegraphics[width=0.45\textwidth]{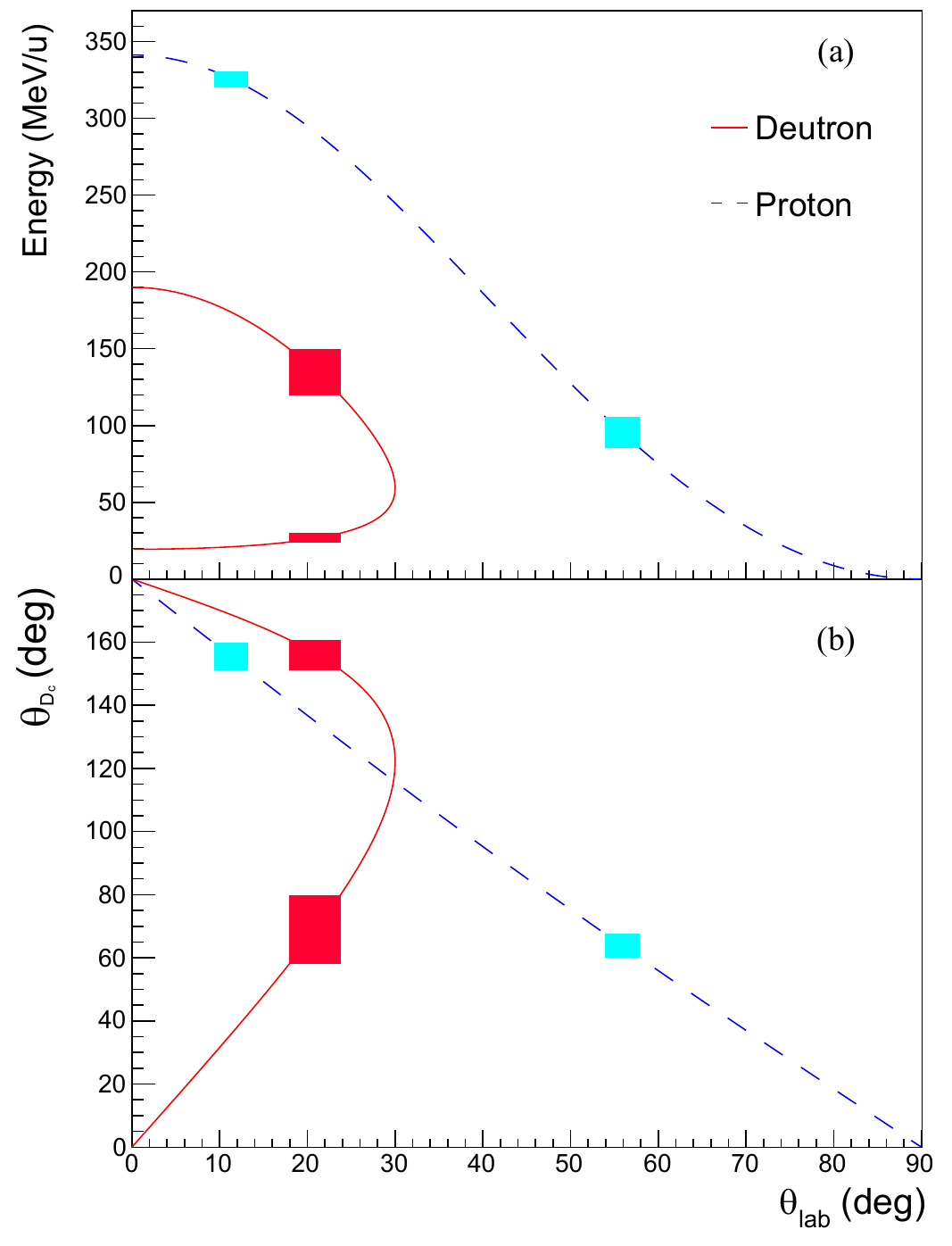}
    \caption{The correlation between the (a) energy and (b) scattering angle of the deuteron and the recoil proton. The location and the angular coverage of the detectors are represented by the colored squares.}
    \label{fig:pol_theta_energy}
\end{figure}

We will perform the experiment with the deuteron beam intensity of $1.6 \times 10^{-3}$ pnA (or $10^7$ pps).Here, pnA stands for particle nano-Ampere, a unit of beam current would be measured if all beam ions were singly charged, and pps stands for particles per second. In order to monitor the beam polarization,  a $1000$ mg/cm$^2$ thick polyethylene (CH2) target is used in the scattering chamber of the polarimeter, which is designed to measure the counts of the elastic scattering deuteron and recoiled proton.  The parameters for the detector placement are listed in Table \ref{tab:detector_placement}. It is designed to detect the recoil protons at two positions $\theta_1$ and $\theta_2$. Figure \ref{fig:pol_theta_energy} shows the coverage of the detectors and the corresponding energy.

\begin{table}[htbp]
    \centering
    \caption{Parameters for the detector placement}
    \begin{tabular}{|c|c|c|c|c|}
            \hline
            Particle & $\theta_{\text{lab}}$ & Distance (mm) & Width  $\theta$(mm) &Width $\phi$(mm)\\
            \hline
            Proton & $\theta_1$ =\ang{55.9} & 600 & 20.0 & 20\\
            Proton & $\theta_2$ = \ang{11.3} & 600 & 20.0 & 20\\
            Deuteron &  \ang{20.87} & 500 & 50 & 40\\
            \hline
    \end{tabular}
    \label{tab:detector_placement}
\end{table}

\begin{figure}[htbp]
        \includegraphics[width=0.45\textwidth]{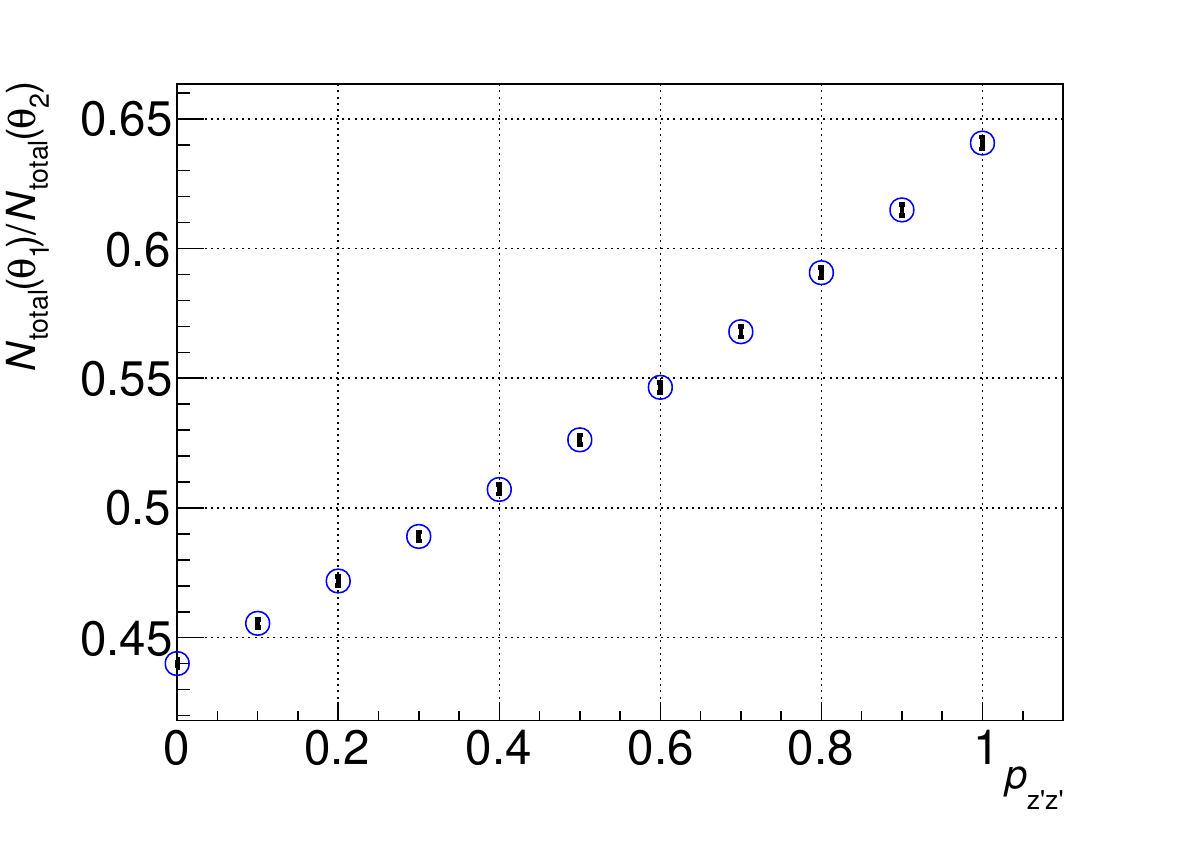}
    \caption{Ratio of the total detector counts at $\theta_1$ and $\theta_2$ during 30 min. The error bars (black lines inside the blue markers)  represent the statistical uncertainty. Due to the large statistics collected in 30 min, these error bars are much smaller than the $\sim10\%$ variation in tensor polarization that can be resolved.}
    \label{fig:pzz_counts}
\end{figure}

\begin{figure}
    \centering
    \includegraphics[width=0.45\textwidth]{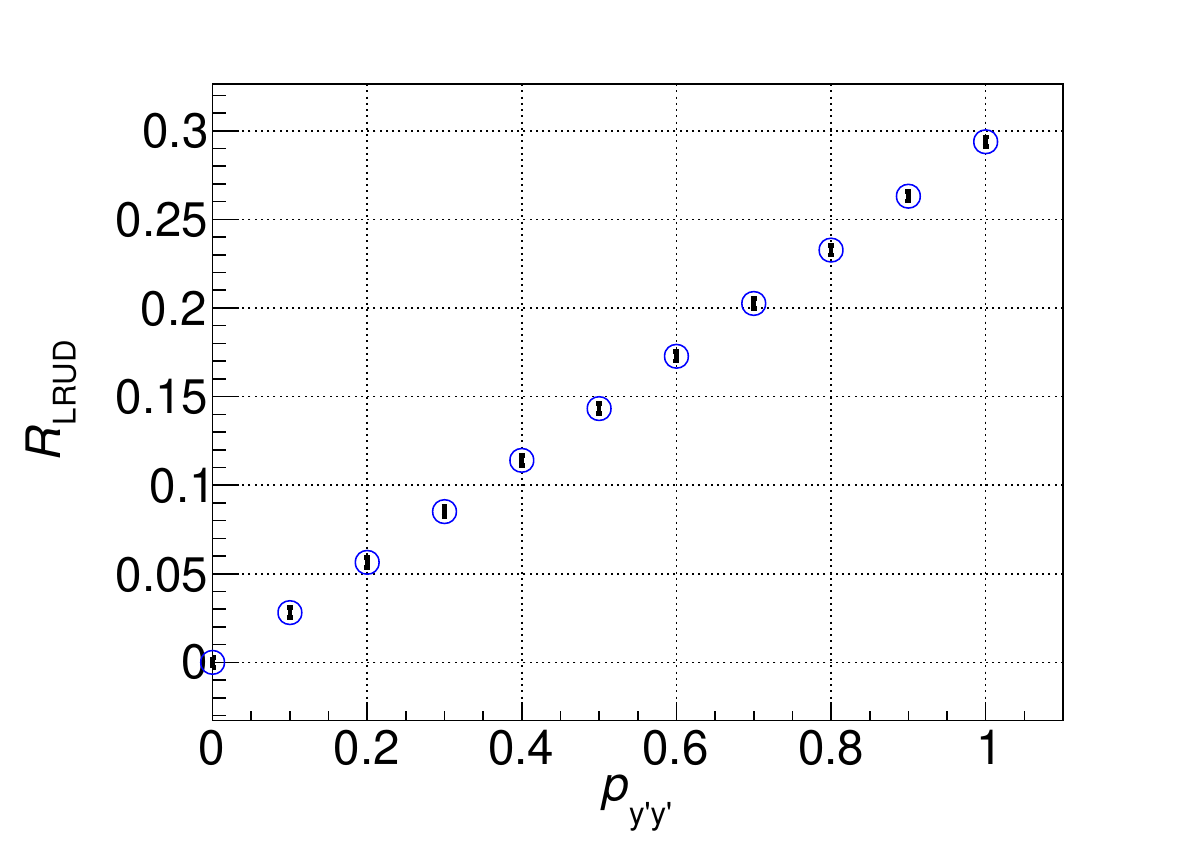}
    \caption{The relationship between $R_{LRUD}$ and the tensor polarization $p_{y'y'}$ during 30 min. The error bars (black lines inside the blue markers) represent the statistical uncertainty.Thanks to the high statistics over the 30 min acquisition, these uncertainties are much smaller than the change caused by $\sim10\%$ polarization variation.}
    \label{fig:pyy_counts}
\end{figure}

Under the given beam intensity, the total counts of the four detectors placed at $\theta_1$ within 30 minutes is about $10^5$ events. Figure \ref{fig:pzz_counts} illustrates the ratio of the detector counts at two different angles as a function of $p_{zz}$. This ensures that we can effectively monitor the longitudinal polarization $p_{zz}$. Similarly, following the previously proposed observable $R_{LRUD}$, its variation as a function of $p_{yy}$ is shown as Figure \ref{fig:pyy_counts}.

The simulation results indicate that the tensor polarization can be monitored with a precision of 10\%, which meets our experimental requirements. Importantly, the data points in Figures~\ref{fig:pzz_counts} and~\ref{fig:pyy_counts} include statistical error bars. The uncertainty at each point is substantially smaller than the typical 10\% variation in polarization that we aim to resolve. This demonstrates that the statistical fluctuations would not mask a 10\% change in tensor polarization. Therefore, the monitoring method is robust and can be reliably applied to similar experiments involving tensor polarization.

\section{Event Generator}
\label{model_desc}
The scattering of polarized deuteron on heavy target is simulated by the ImQMD model, which is an extended version of QMD model suited for the simulation of heavy ion collisions as well as nucleon-induced reactions at intermediate energies. The nucleon is treated as a Gaussian wave packet and follows the Hamiltonian equation of motion in the mean field.  The nuclear potential energy functional including Skyrme potential energy density is written in
 
 \begin{align}\label{Vloc}
 V_{\rm loc}=&\frac{\alpha}{2}\frac{\rho ^{2}}{\rho _{0}}+\frac{\beta }{\eta +1}%
 \frac{\rho ^{\eta +1}}{\rho _{0}^{\eta }}
 +\frac{g_{\rm sur}}{2\rho _{0}}\left(\nabla \rho \right)^{2}
 +\frac{g_{\rm sur,iso}}{\rho_{0}}[\nabla(\rho_{\rm n}-\rho_{\rm p})]^{2}\nonumber\\
 &+g_{\rho\tau}\frac{\rho^{8/3}}{\rho_{0}^{5/3}}+(A\rho+B\rho^{\gamma}+C\rho^{5/3})\delta^{2}\rho,
 \end{align}
 where the isospin asymmetry is $\delta = (\rho_{\rm n} -\rho_{\rm p})/(\rho_{\rm n} +\rho_{\rm p})$, here $\rho_{\rm n}$ and $\rho_{\rm p}$ are the neutron and proton densities \cite{ou2008dynamical}. All parameters in Eq. \ref{Vloc} can be derived from the standard Skyrme interaction parameters. Particularly, the last term of the above equation \ref{Vloc} is replaced by  $\textstyle{C_{\rm s,p}\over 2}\left( \textstyle{\rho \over \rho_0} \right)^{\gamma}\delta^{2}\rho$ to mimic the density dependence of nuclear symmetry energy as density $E_{\rm sym}(\rho)$, while the isoscale part of nuclear potential is unchanged.  With this formulation, the $\gamma$ parameter can characterize the stiffness of $E_{\rm sym}(\rho)$, which  is written as the sum of the kinetic term and the potential term

    \begin{align}
       E_{\rm sym}(\rho)=\frac{C_{\rm s,k}}{2} \left(\frac{\rho}{\rho_0} \right)^{2/3}+ \frac{C_{\rm s,p}}{2} \left(\frac{\rho}{\rho_0} \right)^{\gamma},
       \label{Esymwithgamma}
    \end{align}
where $C_{\rm s,k}$ and $C_{\rm s,p}$ are the kinetic and potential energy parameter, respectively.

\section{Simulation of the detection at SAMURAI}

\label{sec:AngDistribution}
\subsection{SAMURAI Magnet}
The SAMURAI (Superconducting Analyser for Multi-particles from RIKEN) spectrometer, constructed at the RIKEN RI Beam Factory (RIBF), is a large-acceptance multiparticle spectrometer designed to conduct kinematically complete experiments, including the invariant-mass spectroscopy of particle-unbound states in exotic nuclei \cite{Kobayashi2013}. A key component of this advanced system is a superconducting dipole magnet, which, in conjunction with other detectors, enables the coincident detection of heavy fragments and projectile-rapidity nucleons. The magnet has a large bending angle of about $60^\circ$ and a large effective gap of 80 cm. With a maximum field of 3.1~T (field integral 7.1~T$\cdot$m), it is highly capable of analyzing the momentum of heavy fragments \cite{Kobayashi2013}. The different angles are utilized for specific experimental configurations. For example, a $0^\circ$ setup is often used for the detection of positive or negative pions, while a $30^\circ$ angle is the de facto standard configuration for invariant-mass spectroscopy experiments. A $90^\circ$ setting is used for heavy-ion and proton coincidence measurements.

The entire spectrometer system, which also includes beam line detectors, heavy fragment detectors, and a vacuum system \cite{SHIMIZU2013739}, underwent its first commissioning in March 2012. During this commissioning period, the SAMURAI spectrometer demonstrated a rigidity resolution of approximately 1/1500 for RI beams with rigidities of up to 2.4 GeV/c, a testament to the high-performance capabilities of its magnetic component \cite{Kobayashi2013}. The system's ability to perform precise measurements is further enhanced by its detectors for both neutrons and charged particles. Projectile-rapidity nucleons, particularly neutrons, are detected using the NEBULA array.

\subsection{Target and Detector setup}

\begin{figure}
    \centering
    \includegraphics[width=0.45\textwidth]{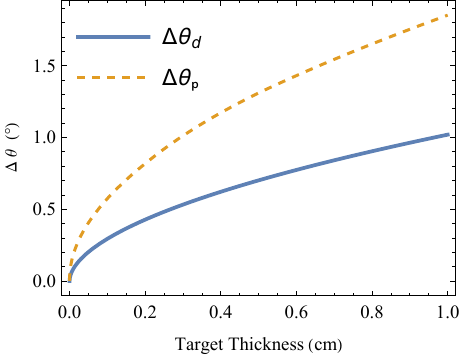}
    \caption{The angular straggling of the target. The values shown are the expectation values calculated by \texttt{lisecute++}.}
    \label{fig:target_angle_straggling}
\end{figure}

We first consider the target. The target thickness is a crucial parameter in the experiment, as it directly affects the detection efficiency and the angular straggling of the particles. A thicker target can increase the number of interactions, but it also increases the angular straggling, which can complicate the momentum reconstruction. Therefore, one seeks to find a balance between the target thickness and the detection efficiency.
 As  Figure \ref{fig:target_angle_straggling} shows, the angular straggling is about $0.5^\circ$ for deutron and $1^\circ$ for proton for a 3 mm thick target. This means that the momentum direction of the detected particles will be affected by this angular straggling, which can lead to a significant uncertainty in the momentum reconstruction. Therefore, we need to carefully consider the target thickness and its impact on the momentum reconstruction.

\begin{figure}
    \centering
    \includegraphics[width=0.45\textwidth]{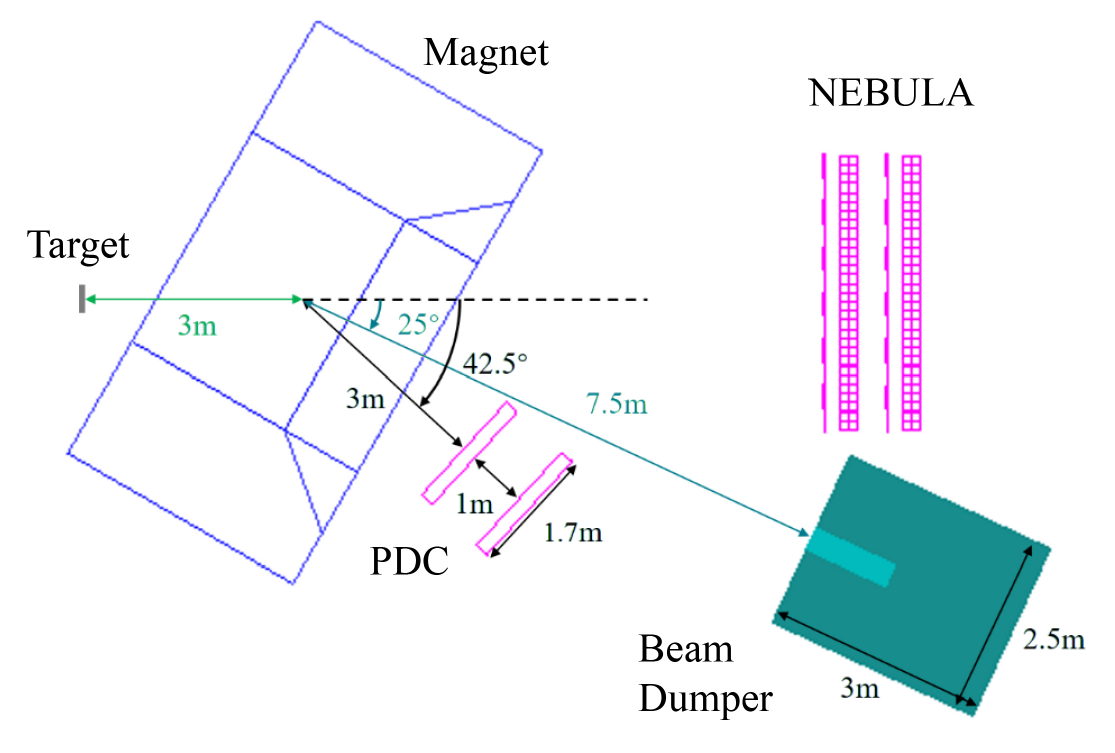}
    \caption{The layout of the particle detector. This configuration is designed to detect the neutron and proton from the deuteron breakup. The target is placed at 3 meters to the field center on the upstream side. Other detectors are shown in the Figure. \ref{fig:exp_plan}.}
    \label{fig:detector_layout}
\end{figure}

Figure \ref{fig:detector_layout} presents the layout of the experiment installed on the SAMURAI terminal. The target is placed at 3 meters to the field center on the upstream side. Neutrons are detected by the NEBULA \cite{NAKAMURA2016156} wall—a large-area neutron detector composed of stacked plastic-scintillator bars (each $\sim$12\,cm $\times$ 12\,cm $\times$ 180\,cm) with PMTs on both ends (and preceding VETO counters)—providing time-of-flight and position information over a wide acceptance. The neutrons are measured by the neutron wall (NEBULA), spanning an acceptance angle of $\pm 12^\circ$ with respect to the current target position, since our target is closer to the NEBULA than the default set \cite{SAMURAI_TAC05}. The central magnetic field strength is about 1.2 Tesla. To achieve a better acceptance angle for protons, the magnet is configured at a $30^\circ$, and the proton drift chamber (PDC) is placed at $42.5^\circ$. The PDC consists of two downstream drift chambers using a cathode-readout method (with U, V, X cathode orientations) to provide high-resolution tracking of projectile-rapidity protons. The path length from the field center to the incident surface of PDC is 3 meters. The two PDCs cover the protons from the deuteron breakup emitted in the horizontal direction within around $\pm 12^\circ$ with respect to the NEBULA.
 
  In this experiment, we are interested in events where the deuteron breaks up into a neutron and a proton while the target nucleus remains intact. In order to suppress the background containing the fragmentation of the target, an active veto detector (VD) consisting of 32 plastic scintillator is placed surrounding the target. The trigger condition requires the coincidence of the NEBULA and SBT, vetoed by the VD.  The three-fold coincidence can suppressed the background greatly.

\subsection{Beam dumper} 

During the beam experiment, intense neutron backgrounds are expected in the NEBULA detector originating from the unreacted beam hitting the environmental material. Thus, to absorb the neutron background, a beam dumper is designed using water tank with dimensions of $2.5 \times 2.5 \times 3~{\rm m^3}$.  An entrance in 34 × 38 × 120 ${\rm cm^3}$ is carved in the front to allow the beam to enter. It is placed at a $25^{\circ}$  and 7.5 m away from the center of the magnet.

\subsection{Observation of the IVR effect in the current experimental design}

In this section, the simulation results of the feasibility to measure the IVR effect are presented. Using the ImQMD event generator, the $\vec{\mathrm{d}} +\mathrm{A}$ scattering at 190 MeV/u are simulated with the impact parameter ranging from 5 to 10 fm.   In the experimental analysis, this range of impact parameters can be selected by applying cuts on observables correlated with $b$, such as the total transverse momentum of the breakup pair, $P_x^{\mathrm{n}}+P_x^{\mathrm{p}}$, and by using centrality detectors like VD. The target "A" is chosen for $^{124}$Sn, $^{112}$Sn and $^{208}$Pb. The parameter $\gamma$ is varying from $\gamma=0.4$ to $0.8$, as defined in Eq.~\ref{Esymwithgamma}, 
(The parameter $\gamma$ characterizes the stiffness of the symmetry energy as shown in Eq. \ref{Esymwithgamma}, and $b$ is the impact parameter.)

The detector response of the $\vec{\mathrm{d}} + \mathrm{A}$ scattering experiment is simulated using the \texttt{\texttt{GEANT4}} toolkit. All detectors are modeled at the individual sub-detector level, allowing precise extraction of both energy deposition and geometric position of each hit. The signals are then smeared according to the detector resolutions. Notably, although the reaction plane , determined  by the incident and scattered particle momenta, is predefined in the simulation, it is not known \textit{a priori} in actual experiments. Therefore, the reaction plane must be reconstructed for both longitudinally and vertically polarized states, which is approximated by $\left(\vec{P}^{\mathrm{n}}+\vec{P}^{\mathrm{p}}\right) \times\hat{z}$.\footnote{The reference frame is built as the Figure \ref{fig:frame}.}

According to the \texttt{GEANT4} simulations using the SAMURAI setup,  totally 2.16M
breakup events  with stiffness parameter $\gamma = 0.6$ 
is recorded  among 10 M deuterons with the impact parameter $b \le 9$ fm, indicating a cross section $\sigma = 550$ mb.

Among these breakup events, 100K events are detected, indicating an overall detection efficiency of  $\eta_{\rm det} = 4.65\%$.

Since the nuclei number density of $^{124}$Sn is $n = N_A \times \rho / M_{\rm Sn} = 3.685 \times 10^{22} \text{cm}^{-3}$,
the total detection number is

\begin{align}
N = n d I \sigma t \eta_{\rm det} = 3.4 \, (d/\text{mm}) (I/\text{pps}) (t/\text{h}),
\end{align}
where $d$ is the target thickness in mm, $I$ is the beam intensity in particles per second (pps) and $t$ is the beam time in hour. 

If we have a target
thickness $d = 3$ mm, beam intensity $I = 10^7$ pps and beam time $t = 16$ h, we will collect about $1.6 \times 10^8$ events.

\begin{figure}[hbt]
    \centering
    \includegraphics[width=0.45\textwidth]{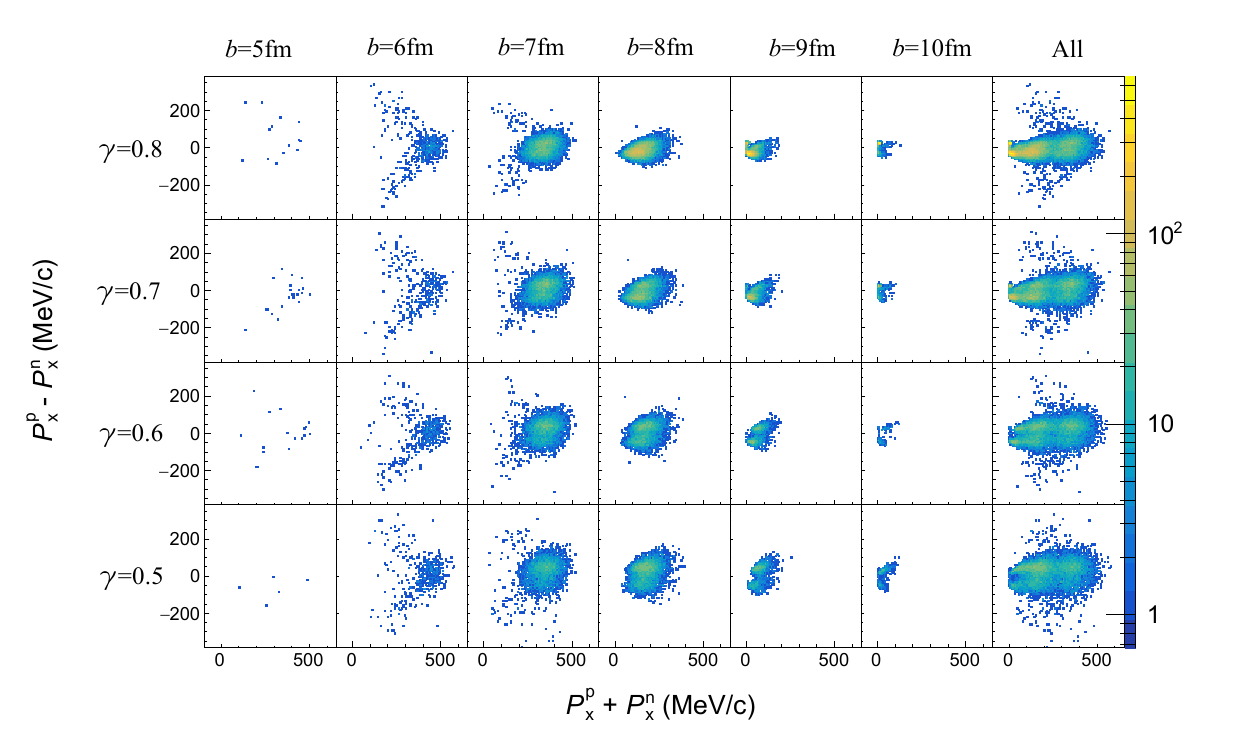}
    \caption{The $P_x$ distribution of the breakup neutrons and protons in the reaction plane. The incident particle is deuteron with $z$ polarization. The abscissa represents the total momentum of the proton and neutron on $x$-axis, which measures the impact parameter $b$ approximately. The ordinate represents the momentum difference between the proton and neutron on $x$-axis, reflecting the magnitude of the isovector force.}
    \label{fig:p_xdistribution}
\end{figure}

In the following,  we describe the simulations conducted to investigate the isovector reorientation effect and the selected observables used to quantify this phenomenon. Additionally, we analyze how these observables vary with the symmetry energy parameter, providing insights into the sensitivity of the observable to the underlying nuclear symmetry energy.

Figure \ref{fig:p_xdistribution} presents the distribution of the momentum of the breakup neutrons and protons in the reaction plane ( along $x$-axis) with different $\gamma$ values. The abscissa is the total momentum of the proton and neutron on the reaction plane $P_x^{\mathrm{n}}+P_x^{\mathrm{p}}$, which measures approximately the impact parameter $b$. Here $P_{x}^{\mathrm{p}}$ and $P_{x}^{\mathrm{n}}$ are the x-momentum of the proton and neutron, respectively.  As one inspects the plots from left to right, when $b<7$ fm, the breakup events is rare because the target is disintegrated.  When the $b$ increases beyond 7 fm, the elastic breakup cross section increases, and the deflection of the deuteron described by $P_x^{\mathrm{n}}+P_x^{\mathrm{p}}$  decreases with $b$. The ordinate represents the momentum difference projected on the reaction plane between the proton and neutron $P_x^{\mathrm{p}}-P_x^{\mathrm{n}}$, reflecting the total effect of the isovector force and the Coulomb force acting on the neutron and the proton. With the $E_{\rm sym}(\rho)$ becomes stiffer by increasing $\gamma$ parameter,  the distribution of momentum difference becomes less populated at the side of $P_x^{\mathrm{p}}-P_x^{\mathrm{n}}>0$.

\begin{figure}
    \centering
    \includegraphics[width=0.45\textwidth]{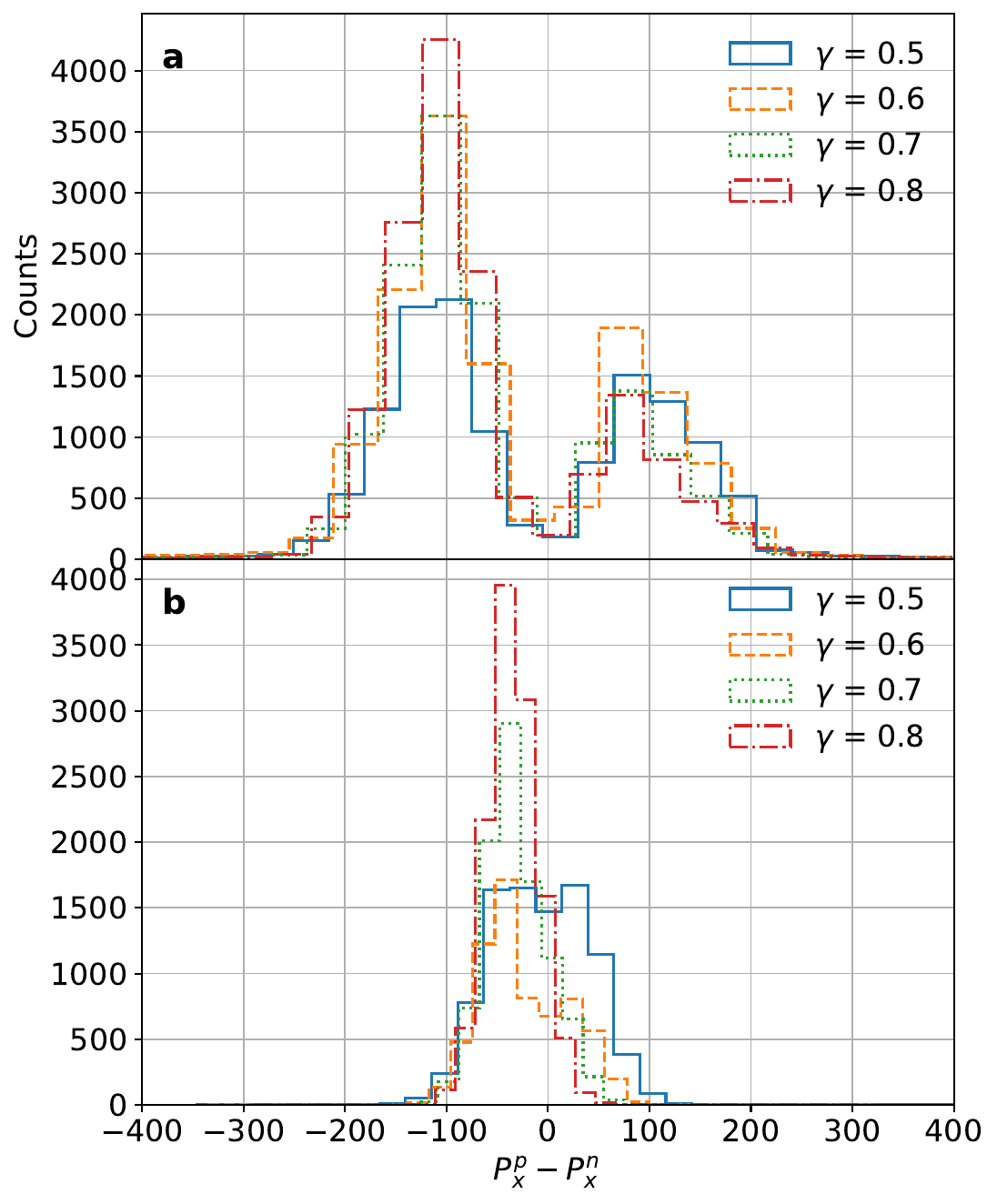}
    \caption{Distribution of $P_{x}^{\mathrm{p}}-P_{x}^{\mathrm{n}}$ (a) for a $y'$-tensor polarized deuteron beam on $^{208}$Pb  and (b) for a $z'$-tensor polarized deuteron beam . The asymmetry in these distributions directly reflects the isovector reorientation effect and its sensitivity to the symmetry energy parameter $\gamma$.}
    \label{fig:Pxp_minus_Pxn_distribution}
\end{figure}

To quantify the isovector reorientation effect, we introduce an observable $R = \frac{N(P_{x}^{\mathrm{p}}>P_{x}^{\mathrm{n}})}{N(P_{x}^{\mathrm{p}}<P_{x}^{\mathrm{n}})}$, representing the ratio of events where the proton's $x$-momentum exceeds that of the neutron to those with the opposite ordering. Figure \ref{fig:Pxp_minus_Pxn_distribution} (a) and (b) present the distribution of $P_{x}^{\mathrm{p}}-P_{x}^{\mathrm{n}}$ for $z'$ and $y'$ tensor polarized deuteron beam on $^{208}$Pb. The observed asymmetry in the distribution indicates that the proton and neutron experience different impulses in the reaction plane, which directly reflects the isovector reorientation effect. Notably, this effect exhibits strong sensitivity to the symmetry energy parameter $\gamma$.

\begin{figure}
    \centering
    \includegraphics[width=0.45\textwidth]{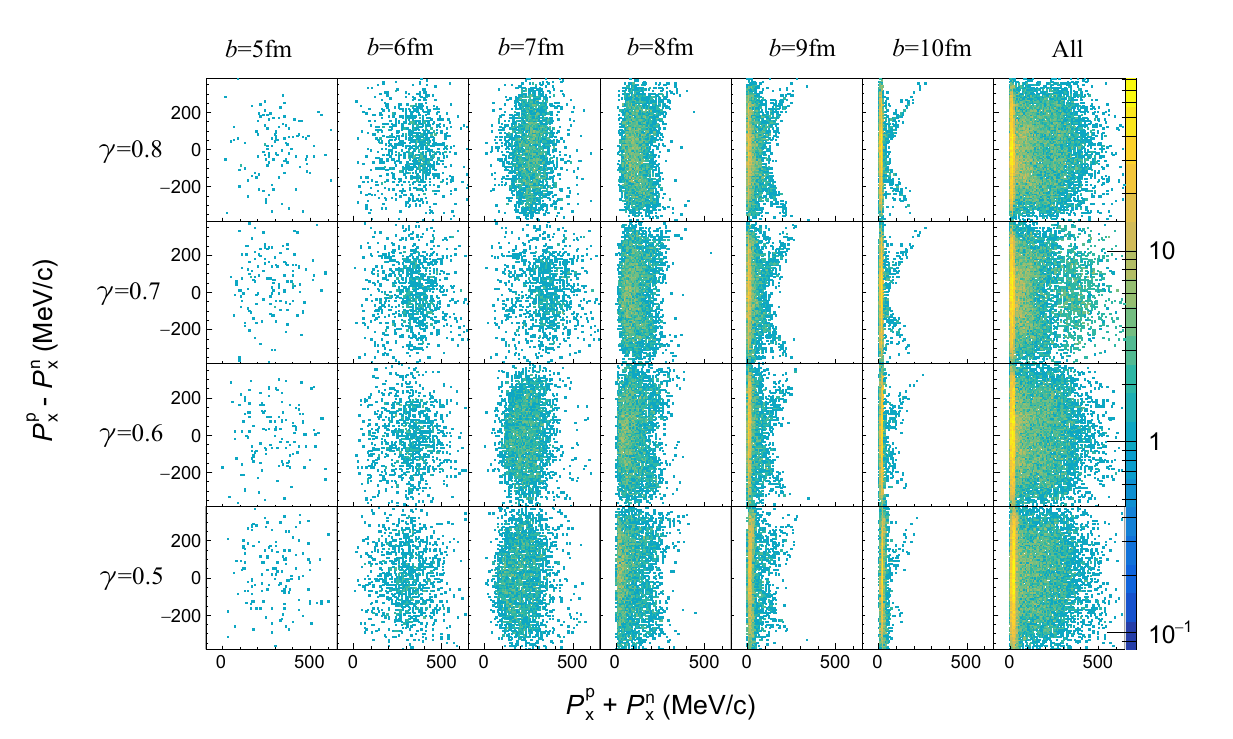}
    \caption{The scattering plot of  $P_{x}^{\mathrm{p}}-P_{x}^{\mathrm{n}}$ versus $P_{x}^{\mathrm{p}}+P_{x}^{\mathrm{n}}$ of the breakup neutrons and protons in the reaction plane. The incident particle is unpolarized deuteron. }
    \label{fig:px_unpolar}
\end{figure}

Figure \ref{fig:px_unpolar} further presents the two-dimensional distribution of $P_{x}^{\mathrm{p}}-P_{x}^{\mathrm{n}}$ versus $P_{x}^{\mathrm{p}}+P_{x}^{\mathrm{n}}$ but for the unpolarized deuteron beam. As expected, although the abscissa $P_{x}^{\mathrm{p}}+P_{x}^{\mathrm{n}}$ exhibits the similar dependence on the impact parameter $b$, the difference of proton and neutron momentum on the reaction plane $P_{x}^{\mathrm{p}}-P_{x}^{\mathrm{n}}$ exhibits left and right symmetry with respect to $P_{x}^{\mathrm{p}}-P_{x}^{\mathrm{n}}=0$ and show no dependence on the stiffness parameter of $\gamma$. 

\begin{figure}
    \centering
    \includegraphics[width=0.45\textwidth]{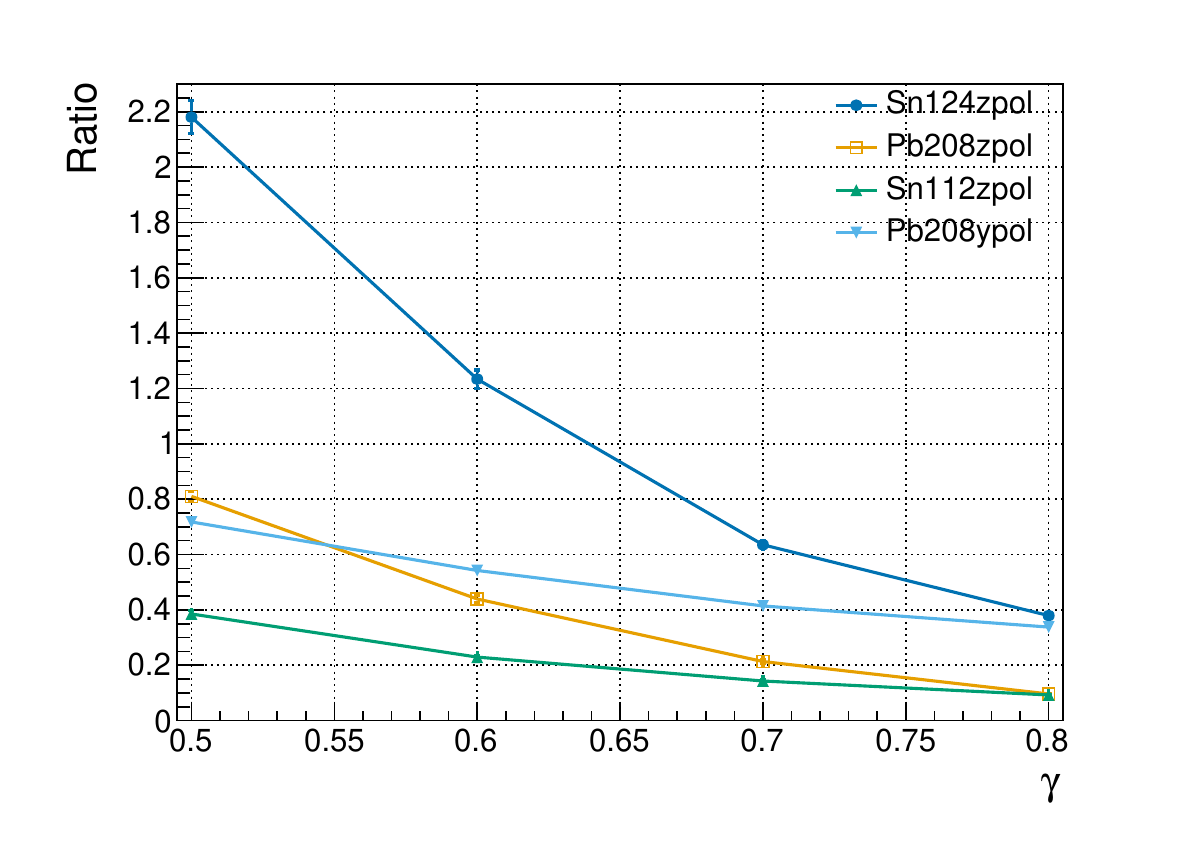}
    \caption{The $R$ observable as a function of the $\gamma$ parameter.}
    \label{fig:IVR_effect}
\end{figure}

Finally, Fig. \ref{fig:IVR_effect} summarizes the dependence of $R$ on the $\gamma$ parameter of the polarized deuteron beam on different target. Some interesting features are visible. First, in the case of $z'$ polarization ($p_{zz}$), it is shown that the sensitivity of $R$ on $\gamma$ is much more pronounced in $^{124}$Sn target than in  $^{208}$Pb because the Coulomb effect is bigger in the latter and counteracts partially against the isospin effect. Next,  the sensitivity is enhanced in the neutron rich target $^{124}$Sn in comparison to that in the less neutron rich target  $^{112}$Sn. Last, the dependence of $R$ on $\gamma$ is stronger with the  $z'$ polarized deuteron beam than that with $y'$ polarization. In all cases, the results demonstrates that the observable $R$ depends on $\gamma$ with significant sensitivity, making it a robust and sensitive probe of the isovector reorientation effect. Thus, the observable  $R$ can serve as a clean probe to constrain the stiffness of the symmetry energy $E_{\mathrm{sym}}(\rho)$ in nuclear matter.

\section{Conclusion}
\label{sec:Conclusion}

The experiment of polarized deuteron scattering on various heavy target aims to investigate the isovector reorientation (IVR) effect and its inference of the stiffness of nuclear symmetry energy at sub-saturation density. The experiment is proposed to run at the SAMURAI spectrometer at RIKEN, utilizing a polarized deuteron beam with an energy of 190 MeV/u. In this paper, we demonstrate the method to monitor the $p_{z'z'}$ polarization and verify the effectiveness of $p_{y'y'}$ polarization monitoring proposed by previous studies. We have designed the detector setup and optimized the target position to maximize the detection efficiency for both neutrons and protons. The responses of the detectors to these particles were thoroughly analyzed, along with simulations of the background and the beam dumper to minimize the interference from the background. It is found that, the IVR effect, measured by the observable $R$, is more pronounced in the neutron-rich target $^{124}$Sn than in the neutron-deficient $^{112}$Sn, and also more pronounced in $^{124}$Sn than in $^{208}$Pb. Moreover, the $z'$-polarized beam is more favored than the $y'$-polarized beam.  In all cases, the IVR effect can be clearly observed in the simulations.  The simulation predictions demonstrate the feasibility of the experiment.

\section*{Acknowledgments}  

This work is supported by the Ministry of Science and Technology of China under Grant Nos. 2022YFE0103400  and by the Tsinghua University Initiative Scientific Research Program. Support from the Open Project Program of State Key Laboratory of Theoretical Physics, Institute of Theoretical Physics, Chinese Academy of Sciences, China (No. Y4KF041CJ1) is also gratefully acknowledged. 
We are grateful to the SAMURAI collaboration for valuable discussions and assistance during the simulation studies. 

\section*{Declarations}
\textbf{Conflicts of Interest}
On behalf of all authors, the corresponding authors state that there is no conflict of interest.

\bibliography{ref_tbt}

\end{document}